# Incremental auxetic response of composite lattices under isotropic prestress

Ada Amendola[a], Fernando Fraternali[a]

[a]*Department of Civil Engineering, University of Salerno, Via Giovanni Paolo II, 132, 84084 Fisciano (SA), Italy*

**Abstract**

   This work studies the constitutive response of two- and three-dimensional lattice materials subject to isotropic prestress. The unit cell of the examined lattices is formed by an arbitrary number of junctions attached to a junction. Analytic formulae for second-order elastic constants and elastic stiffness coefficients of stretching-dominated lattices are provided. In addition, numerical results for the incremental elastic moduli of composite lattices equipped with hard and soft rods are presented. The given results highlight that isotropically pre-tensioned lattices may exhibit marked variations of the elastic stiffness moduli in the prestressed state, over the values competing to the stress-free configuration. This study also discloses that prestressed lattices may feature incremental auxetic response, when composite architectures suitably combining hard and soft materials are employed for their fabrication.

*Keywords:* lattice materials; stretching-dominated response; prestress; elastic moduli; auxetic response.

## 1  Introduction

   Recent research in the area of lattice mechanics has revealed that suitably designed, micro- and nano-lattice materials are able to achieve extreme properties over more than several orders of magnitude in density, being able to fill holes in the current material property charts, through an optimal control of material and space [1]-[19]. Stretching-dominated lattices can achieve extremely high stiffness-to-density ratios [6]-[19], while bending-dominated lattices generally exhibit more compliant but more recoverable response under large strains [3, 4, 15]. Hierarchical architectures and structures equipped with hollow tubes have been employed to combine high strength and high recoverability [6, 7, 14, 17, 18], since standard stretching-dominated lattice materials are typically affected by failure under buckling, which may reduce strength and cyclability of the material [7, 13, 14]. Increasing attention are also receiving the so-called isogrid and anisogrid lightweight structures [20]-[23]. A special class of lattice materials with extremal response is that of *auxetic lattices*, which are endowed with peculiar microstructures (e.g., honeycomb architectures with re-entrant corners) that allow the material to exhibit strains of equal signs both in the direction of the applied load, and in the transverse direction (i.e., negative Poisson's ratios) [24]. It has been show that auxetic structures are well suited for the manufacturing of impact- and vibration-resistant materials, in biomedical applications, and for the fabrication of innovative fibre-reinforced composites, just to mention few relevant examples [25]-[28].
   The effects of initial stresses on the incremental constitutive response of elastic solids have been longly investigated in the literature (refer, e.g., to [29]-[36] and references therein). Initial stresses may refer to the state of stress of a body that has undergone a previous story of finite elastic deformations (*prestressed body*), and/or to self-equilibrated (or *residual*) stresses arising in a body in absence of external loads, due to manufacturing processes, growth processes, hygrothermal effects, etc. [35]. Such stresses are diffusely present in biological structures, like arteries and soft tissues [34], and their action may significantly affect the constitutive response and the propagation of mechanical waves in elastic materials [35, 36]. The effects of geometric stiffness terms due to large displacements and initial stresses in tensegrity structures and other prestressed lattices are diffusely analyzed in [37, 38, 39] and references therein. The relevant role played by such effects on the incremental response of prestressed lattices structures is nowadays well recognized in the literature, but most of the available studies are focused on finite size structures, rather than the continuous response of infinite networks at the mesoscale (refer, e.g., to [38] for an overview).

Additive manufacturing has become the most common technique for fabricating periodic lattices materials at different scales (refer, e.g., to [40, 41] and references therein). Several fabrication methods have been proposed in this field, with resolution ranging from the centimeter- to the nanometer-scale. Worth mentioning here are: fused deposition modeling (FDM); polyjet 3D printing technologies; electron beam melting; x-ray lithography; deep ultraviolet lithography; soft lithography; two-photon polymerization; atomic layer deposition; and projection micro-stereolithography, among other available methods [40]-[43]. The application of prestress states in lattices fabricated through additive manufacturing, however, is an open issue at present.

The present study investigates the incremental elastic response of 2d and 3d stretching-dominated lattice materials, which are subject to a homogeneous state of isotropic prestress induced by a finite deformation from the natural (or *stress free*) state. The targeted lattice materials may consist of sandwich structures with truss cores and restraining facesheets [44], which are subject to initial pre-tensioning through suitable actuation devices [37, 45]. We study the kinematics, equilibrium equations and incremental elastic energy of the isotropically prestressed lattices (Sects. 2-4), assuming that their unit cell is formed by an arbitrary number of rods attached to a junction. Next, we obtain analytic expressions of second-order elastic constants and elastic stiffness coefficients [29]-[33] (Sect. 4). The adopted approach to the elastic moduli of isotropically prestressed lattices reduces to that given in [33] for statistical mechanics systems at zero temperature. We apply the obtained formulae for the elastic constants to noticeable 2d and 3d examples of lattice materials under isotropic prestress (Sect. 5), which are different from those studied in [38]. Such a study is focused on composite lattices that can be additively manufactured, and are formed by rods made of either 'hard' and/or 'soft' FDM materials [46, 47]. The given numerical results reveal novel features of the mechanical response of isotropically pre-tensioned, hard/soft composite lattices, which consist of their ability to exhibit significant variations of the elastic moduli from the prestressed state, as compared to the moduli characterizing the stress-free configuration. In particular, the examined lattice materials are found capable to show marked auxetic incremental response from the prestressed state. Concluding remarks and directions for future work are presented in Sect. 6.

## 2 Kinematics of a prestressed lattice

Let us consider an arbitrary lattice whose unit cell features $Z$ rods attached to a a junction. On introducing a Cartesian reference frame with the origin at the position of the junction in the prestressed configuration, we consider a homogeneous (or affine) incremental deformation of the unit cell defined through the following deformation map

$$\mathbf{r}_i = \mathbf{F} \mathring{\mathbf{r}}_i \quad (1)$$

where $\mathring{r}_i$ is the position vector of the end-point of the $i$-th rod in the prestressed configuration ($i = 1, ..., Z$), $r_i$ is the position vector of the same point in the deformed configuration, and $F$ is a constant deformation gradient. Denoted the norms of $r_i$ and $\mathring{r}_i$ by $r_i$ and $\mathring{r}_i$, respectively, we easily obtain from Eqn.(1) the well known formula

$$r_i^2 = \mathring{r}_i^2 + 2\mathbf{E}\,\mathring{\mathbf{r}}_i \cdot \mathring{\mathbf{r}}_i \quad (2)$$

where

$$E = \tfrac{1}{2}(F^T F - I) \quad (3)$$

is the Green-Saint Venant finite strain tensor [48, 49]. Here and in what follows, the apex $T$ denotes the transposition symbol, and $I$ denotes the identity tensor. Let now $\varepsilon$ denote the norm of the displacement gradient $H = F - I$. Making use of Taylor's formula, we obtain from Eqn.(2) the result

$$r_i - \mathring{r}_i = \mathring{r}_i + \mathbf{E}\mathring{\mathbf{e}}_i \cdot \mathring{\mathbf{e}}_i - \frac{\mathring{r}_i}{2}\left(\mathbf{E}\mathring{\mathbf{e}}_i \cdot \mathring{\mathbf{e}}_i\right)^2 + o(\varepsilon^2) \quad (4)$$

where $o(\alpha)$ denotes a scalar function that approaches zero faster than $\alpha$, and $\mathring{\mathbf{e}}_i$ denotes the unit vector in the direction of $\mathring{\mathbf{r}}_i$.

## 3 Incremental equilibrium equations in the prestressed state

We model each rod of the lattice as a linear spring obeying the following constitutive law
$$f_i = k_i(r_i - \bar{r}_i) \quad \mathbf{e}_i \tag{5}$$
In Eqn.(5), $f_i$ is the axial force carried by the generic rod (say, the $i$-th one), $\bar{r}_i$ is the length of such a rod in the natural state (*rest length*), $\mathbf{e}_i$ is the unit vector in the direction of the rod in the deformed configuration, and $k_i$ is a constant (i.e., deformation independent) stiffness coefficient given by
$$k_i = \frac{E_i A_i}{\bar{r}_i} \tag{6}$$
where $A_i$ denotes the cross-section area of the rod. The latter is supposed to remain constant during any lattice deformation.

Our next developments make use of the assumption that the lattice is at equilibrium under zero external forces in correspondence of the inner nodes, both in the prestressed state and in the deformed configuration. Such an assumption implies that the rods follow any mesoscopic deformation of the lattice at the microscopic level by deforming in the pure stretching mode (no bending effects) [9]. The equilibrium equations in the deformed configuration require that at each inner node it results
$$\sum_{i=1}^{Z} \mathbf{f}_i = \sum_{i=1}^{Z} k_i (r_i - \mathring{r}_i + \mathring{r}_i - \bar{r}_i) \mathbf{e}_i = \sum_{i=1}^{Z} \left\{ k_i (r_i - \mathring{r}_i) + k_i(\mathring{r}_i - \bar{r}_i) \right\} \mathbf{e}_i = \mathbf{0} \tag{7}$$
while those in the prestressed state impose
$$\sum_{i=1}^{Z} \mathring{\mathbf{f}}_i = \sum_{i=1}^{Z} k_i (\mathring{r}_i - \bar{r}_i) \mathring{\mathbf{e}}_i = \mathbf{0} \tag{8}$$

Taking to into account Eqn. (8) and retaining terms up to the first-order in $\varepsilon$, we deduce from Eqn. (7) the following incremental equilibrium equations
$$\sum_{i=1}^{Z} k_i (r_i - \mathring{r}_i) \mathring{\mathbf{e}}_i = \mathbf{0} \tag{9}$$

## 4 Elastic constants

The elastic energies competing to the unit cell in the prestressed and deformed configurations are respectively given by
$$\mathcal{E}_0 = \frac{1}{2} \sum_{i=1}^{Z} k_i (\mathring{r}_i - \bar{r}_i)^2 \tag{10}$$
$$\mathcal{E} = \frac{1}{2} \sum_{i=1}^{Z} k_i (r_i - \bar{r}_i)^2 \tag{11}$$
and it is immediate to verify that it results
$$\mathcal{E} - \mathcal{E}_0 = \sum_{i=1}^{Z} \mathring{t}_i (r_i - \mathring{r}_i) + \frac{1}{2} \sum_{i=1}^{Z} k_i (r_i - \mathring{r}_i)^2 \tag{12}$$
where $\mathring{t}_i$ denotes the scalar axial force carried by the $i$-th rod in the prestressed configuration. According to the constitutive equation (5), such a quantity is given by

$$\mathring{t}_i = k_i \left( \mathring{r}_i - \bar{r}_i \right) \tag{13}$$

By making use of Eqn. (4) into Eqn. (12), we obtain

$$\frac{\mathcal{E} - \mathcal{E}_0}{V_0} = \frac{1}{V_0} \sum_{i=1}^{Z} \mathring{t}_i \mathring{r}_i E_{\alpha\beta} \mathring{e}_{i_\alpha} \mathring{e}_{i_\beta} + \frac{1}{2V_0} \sum_{i=1}^{Z} \left( k_i \mathring{r}_i - \mathring{t}_i \mathring{r}_i \right) E_{\alpha\beta} E_{\lambda\mu} \mathring{e}_{i_\alpha} \mathring{e}_{i_\beta} \mathring{e}_{i_\lambda} \mathring{e}_{i_\mu} + o(\varepsilon^2) \tag{14}$$

where $V_0$, indicates the volume of the unit cell in the prestressed state. In Eqn. (14), we have denoted the Cartesian components of $E$ and $\circ e_i$ by $E_{\alpha\beta}$ and $\circ e_{i_\alpha}$ ($\alpha, \beta = 1, .., d$), respectively, and we have made use of the summation convention on Greek indices.

The structure of Eqn.(14) leads us to recognize that, at the mesoscale, the prestressed lattice exhibit the following homogeneized *Cauchy stress* $\circ \sigma_{\alpha\beta}$ and the incremental (or tangent) *second-order elastic constants* $\circ C_{\alpha\beta\lambda\mu}$ (cf. [29], par. 4.2)

$$\mathring{\sigma}_{\alpha\beta} = \frac{1}{V_0} \sum_{i=1}^{Z} \mathring{t}_i \mathring{r}_i \mathring{e}_{i_\alpha} \mathring{e}_{i_\beta} \tag{15}$$

$$\mathring{C}_{\alpha\beta\lambda\mu} = \frac{1}{V_0} \sum_{i=1}^{Z} \left( k_i \mathring{r}_i^2 - \mathring{t}_i \mathring{r}_i \right) \mathring{e}_{i_\alpha} \mathring{e}_{i_\beta} \mathring{e}_{i_\lambda} \mathring{e}_{i_\mu} \tag{16}$$

It is easily verified that $\circ \sigma_{\alpha\beta}$ is symmetric in $\alpha, \beta$ and that the elastic constants $\circ C_{\alpha\beta\lambda\mu}$ match the following full symmetry conditions

$$\mathring{C}_{\alpha\beta\lambda\mu} = \mathring{C}_{\beta\alpha\lambda\mu} = \mathring{C}_{\alpha\beta\mu\lambda} = \mathring{C}_{\lambda\mu\alpha\beta} \tag{17}$$

Such constants have full mathematical and thermodynamical meaning, but cannot be measured experimentally (see, e.g., [29, 33]). Elastic constants that instead characterize the experimental response of the lattice material at the mesoscopic level in the prestressed state are the so-called *elastic stiffness coefficients* $\circ c_{\alpha\beta\lambda\mu}$, which are defined through (refer, e.g., to [29, 33])

$$\mathring{c}_{\alpha\beta\lambda\mu} = \mathring{C}_{\alpha\beta\lambda\mu} - \frac{1}{2} \left( 2 \mathring{\sigma}_{\alpha\beta} \delta_{\lambda\mu} - \mathring{\sigma}_{\alpha\lambda} \delta_{\beta\mu} - \mathring{\sigma}_{\alpha\mu} \delta_{\beta\lambda} - \mathring{\sigma}_{\beta\lambda} \delta_{\alpha\mu} \right) \tag{18}$$

where $\delta_{\lambda\mu}$ denotes the Kronecker delta. In the case under consideration it results

$$\mathring{\sigma}_{\alpha\beta} = \delta_{\alpha\beta} \mathring{\sigma} \tag{19}$$

$\circ \sigma$ being the applied isotropic stress. Making use of Eqn.(19) into Eqn.(18), we obtain

$$\mathring{c}_{\alpha\beta\lambda\mu} = \mathring{C}_{\alpha\beta\lambda\mu} - \mathring{\sigma} \left( \delta_{\alpha\beta} \delta_{\lambda\mu} - \delta_{\alpha\lambda} \delta_{\beta\mu} - \delta_{\alpha\mu} \delta_{\beta\lambda} \right) \tag{20}$$

which implies that the elastic stiffness coefficients $\circ c_{\alpha\beta\lambda\mu}$ exhibit full symmetry properties, in the present case. The same result does not hold under more general conditions of initial stress, when such constants exhibit only minor symmetry properties (cf. [29]).

It is worth observing that the elastic stiffness coefficients $\circ c_{\alpha\beta\lambda\mu}$ coincide with the elastic constants $C_{\alpha\beta\lambda\mu}$ that match material stability conditions in statistical mechanics systems at zero temperature [33], for the case under consideration of isotropic prestress.

**5   2D and 3D examples**

The present section is devoted to derive the expressions of the second order and elastic stiffness constants of some noticeable lattice materials under isotropic prestress. The given results show that all the examined lattices exhibit incremental elastic response from the prestressed state characterized by cubic symmetry with material symmetry axes aligned with the unit cell edges. We assume that the unit vectors $\mathring{e}_i$ in the prestressed state of such lattices are aligned with those characterizing the natural state. Under such an assumption, it is easily shown that the equilibrium equations (8) are satisfied for any arbitrary incremental deformation of the examined lattices from the prestessed state (see also [9]). We develop numerical predictions of the elastic moduli of composite lattices that can be 3d printed in hard or soft FDM materials, namely the ABS-M30 by Stratasys, with Young modulus $\hat{E} = 2.2$ GPa, and yield stress $\sigma_y = 30$ MPa [46] (hereafter simply denoted as 'ABS' or 'Hard'), and the NinjaFlex by NinjaTek, a flexible polyurethane material that exhibits Young modulus $\hat{E} = 12$ MPa, and yield stress $\sigma_y = 4$ MPa [47] (hereafter referred to as 'NJF' or 'Soft'). Since compressed lattices may easily fail in buckling (refer, e.g., to [13, 14] and references therein), we focus our attention on systems loaded by isotropic prestress in tension, i.e. equal biaxial or triaxial tensile loads.

### 5.1 Tetrakis-like lattices

We begin by examining a two-dimensional tetrakis-like lattice material that tessellates the plane through the unit cell illustrated in Figure 1. Such a unit cell is equipped with $Z = 8$ rods, which we assume are endowed with the following geometrical and prestress properties

$$\mathring{e}_1, \mathring{e}_5 = [\pm 1, 0]^T, \quad \mathring{e}_2, \mathring{e}_4, \mathring{e}_6, \mathring{e}_8 = \left[\pm \frac{h_1}{\sqrt{h_1^2 + h_2^2}}, \pm \frac{h_2}{\sqrt{h_1^2 + h_2^2}}\right]^T, \quad \mathring{e}_3, \mathring{e}_7 = [0, \pm 1]^T, \tag{21}$$

$$\mathring{r}_1 = \mathring{r}_5 = h_1, \quad \mathring{r}_2 = \mathring{r}_4 = \mathring{r}_6 = \mathring{r}_8 = \frac{\sqrt{h_1^2 + h_2^2}}{2}, \quad \mathring{r}_3 = \mathring{r}_7 = h_2 \tag{22}$$

$$\mathring{k}_1 = \mathring{k}_5 = \frac{\hat{E}_1 A_1}{h_1}, \quad \mathring{k}_2 = \mathring{k}_4 = \mathring{k}_6 = \frac{2 \hat{E}_2 A_2}{\sqrt{h_1^2 + h_2^2}}, \quad \mathring{k}_3 = \mathring{k}_7 = \frac{\hat{E}_3 A_3}{h_2} \tag{23}$$

$$\mathring{t}_1 = \mathring{t}_5 = \frac{1}{2h_1} V_0 \sigma_{01} + \frac{h_2^2 - h_1^2}{2h_1 h_2^2} V_0 \sigma_{02}, \quad \mathring{t}_2 = \mathring{t}_4 = \mathring{t}_6 = \frac{\sqrt{h_1^2 + h_2^2}}{2h_2^2} V_0 \sigma_{02}, \quad \mathring{t}_3 = \mathring{t}_7 = \frac{1}{2h_2} V_0 \sigma_{01} \tag{24}$$

In the above equations, $A_1$, $A_2$ and $A_3$ denote the cross-section areas of horizontal, vertical and diagonal rods (Fig. 1); $\hat{E}_1$, $\hat{E}_2$ and $\hat{E}_3$ denote the Young moduli of the materials forming such rods; and $\sigma_{01}$ and $\sigma_{02}$ indicate two different states of isotropic prestress of the lattice. Throughout the manuscript, we convene to affect with the superscript $\hat{}$ the properties of the materials that form the rods, in oder to distinguish such properties from those relative to the lattice material at the mesoscale. As seen from Eqns. (24), the state of prestress corresponding to $\sigma_{01}$ involves prestress forces only in horizontal and vertical members, while that corresponding to $\sigma_{02}$ involves prestress forces in horizontal and diagonal members.

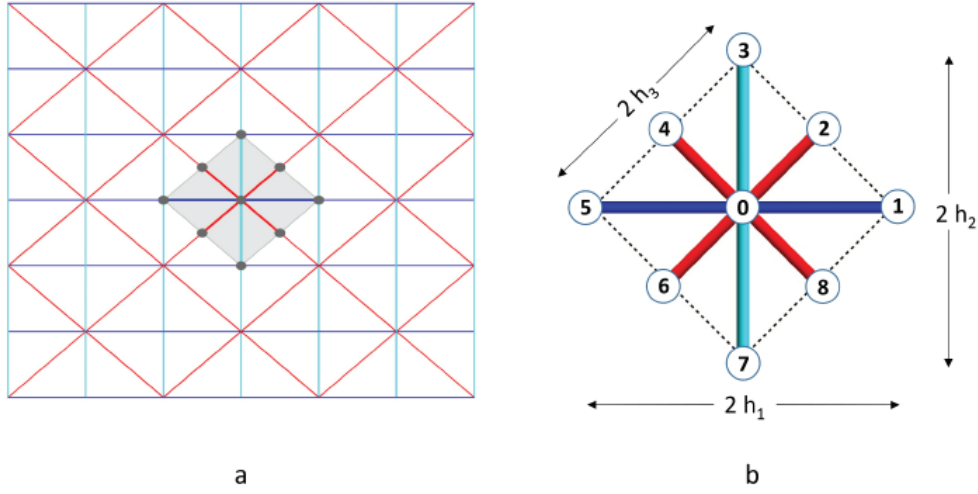

Figure 1: Tetrakis-like lattice material (left), and zoom-in of the elementary unit cell in the prestressed state (right) ($h_3 = \sqrt{h_1^2 + h_2^2}/2$).

### 5.1.1 Second-order elastic constant

Let us make use of Eqns.(21) - (24) into Eqns. (16) to obtain the second-order elastic constants of the lattice material under consideration in the following matrix form (Voigt notation)

$$\mathring{\mathbf{C}} = \begin{bmatrix} \mathring{C}_{11} & \mathring{C}_{12} & 0 \\ \mathring{C}_{12} & \mathring{C}_{22} & 0 \\ 0 & 0 & \mathring{C}_{44} \end{bmatrix}$$

(25)

where

$$\mathring{C}_{11} = \mathring{C}_{1111} = \frac{2 A_2 \hat{E}_2 h_1^4 h_3 + A_1 \hat{E}_1 h_1 16 h_3^4 - h_3^2 \left(h_1^2 \sigma_{01} + h_2^2 \sigma_{02}\right)}{2 h_3^2},$$

$$\mathring{C}_{22} = \mathring{C}_{2222} = \frac{A_3 \hat{E}_3 h_2 8 \sqrt{h_3^6} + A_2 \hat{E}_2 h_2^4 - h_3 \left(h_1^2 \sigma_{01} + h_2^2 \sigma_{02}\right)}{4 \sqrt{h_3^6}},$$

$$\mathring{C}_{12} = \mathring{C}_{1122} = \mathring{C}_{2211} = -\frac{h_1^2 \left(-A_2 \hat{E}_2 h_2^2 h_3 + h_3^2 \sigma_{02}\right)}{4 h_3^4},$$

$$\mathring{C}_{44} = \mathring{C}_{1122} = \mathring{C}_{12}.$$

(26)

### 5.1.2 Elastic stiffness coefficients

The use of Eqns. (26) into Eqns. (20) leads us to following expressions of the matrix of the experimental elastic constants

$$\mathring{c} = \begin{bmatrix} \mathring{c}_{11} & \mathring{c}_{12} & 0 \\ \mathring{c}_{12} & \mathring{c}_{22} & 0 \\ 0 & 0 & \mathring{c}_{44} \end{bmatrix} \quad (27)$$

Where

$$\mathring{c}_{11} = \frac{h_1 \left( A_2 \hat{E}_2 h_1^3 h_3 + A_1 \hat{E}_1 8 h_3^4 + h_1 h_3^2 \sigma_{02} V_0 \right)}{4 h_3^4 V_0},$$

$$\mathring{c}_{12} = \frac{A_2 \hat{E}_2 h_1^2 h_2^2 h_3 - h_3^2 \left[ h_2^2 \sigma_0 + h_1^2 (\sigma_{01} + \sigma_{02}) \right] V_0}{4 h_3^4 V_0},$$

$$\mathring{c}_{22} = \frac{A_2 \hat{E}_2 h_2^4 h_3 + A_3 \hat{E}_3 h_2 8 h_3^4 + h_1^2 h_3^2 \sigma_{02} V_0}{4 h_3^4 V_0},$$

$$\mathring{c}_{44} = \frac{A_2 \hat{E}_2 h_1^2 h_2^2 h_3 + h_3^2 \left( h_1^2 \sigma_{01} + h_2^2 \sigma_0 \right) V_0}{4 h_3^4 V_0}.$$

(28)

### 5.1.3 Elastic stiffness moduli

In order to get handy expressions of the Young moduli, Poisson's ratios and shear modulus of a prestressed tetrakis lattice material, which can be actually measured in experimental tests (*elastic stiffness moduli*), from now on we will focus our attention on a square tetrakis lattice that features $h_1 = h_2 = h$, $h_3 = h\sqrt{2}/2$, $A_3 = A_1$, $A_2 = \sqrt{2} A_1$, and $\hat{E}_3 = \hat{E}_1$ (horizontal and vertical rods made of the same material). In the present case, we refer to the horizontal and vertical rods as members of 'Type 1', and to the diagonal rods as members of 'Type 2'.

By inverting the matrix $\circ c$ given by Eqns. (27)-(28), we obtain the compliance matrix $\circ a = \circ c^{-1}$ and the following expressions of the elastic stiffness moduli of the prestressed lattice

$$E = \frac{1}{\mathring{a}_{11}} = \frac{1}{\mathring{a}_{22}} = \frac{\left( \hat{E}_1 \phi + 4 \sigma_{01} + 8 \sigma_{02} \right) \left[ \left( \hat{E}_1 + \hat{E}_2 \right) \phi - 4 (\sigma_{01} + \sigma_{02}) \right]}{2 \left( 2 \hat{E}_1 + \hat{E}_2 \right) \phi + 8 \sigma_{02}},$$

$$\nu = -\mathring{a}_{12} E_{11} = -\mathring{a}_{21} E_{22} = \frac{\hat{E}_2 \phi - 8 \sigma_{01} - 12 \sigma_{02}}{\left( 2 \hat{E}_1 + \hat{E}_2 \right) \phi + 4 \sigma_{02}},$$

(29)

$$G = \frac{1}{\mathring{a}_{44}} = \mathring{c}_{44} = \frac{1}{8} \left( \hat{E}_2 \phi + 4 \sigma_{02} \right) + \sigma_{01},$$

$$K = \frac{1}{2(\mathring{a}_{11} + \mathring{a}_{12})} = \frac{\mathring{c}_{11} + \mathring{c}_{12}}{2} = \frac{1}{8} \left[ (\hat{E}_1 + \hat{E}_2) \phi - 4 \sigma_{01} - 4 \sigma_{02} \right].$$

In Eqns. (29), $\phi$ denotes the solid volume fraction of the lattice in the prestressed state (defined as the ratio $V_S/V_0$ between the total volume of the rods in the unit cell $V_S$ and the unit cell volume $V_0$), while $E$ denotes the Young modulus, $\nu$ denotes the Poisson's ratio, $G$ denotes the shear modulus, and $K$ denoted the bulk modulus of the cubic-symmetric material, which characterize the incremental experimental response of the material from the prestressed state. When it results $\sigma_{01} = \sigma_{02} = 0$, from Eqns. (29) we deduce the following expressions of the engineering elastic constants in the stress-free state ('stress-free moduli', cf. Biot [31])

$$\begin{aligned}
\bar{E} &= \frac{\hat{E}_1}{2} \frac{(\hat{E}_1 + \hat{E}_2)}{(2\hat{E}_1 + \hat{E}_2)} \phi, \\
\bar{\nu} &= \frac{\hat{E}_2}{2\hat{E}_1 + \hat{E}_2}, \\
\bar{G} &= \frac{\hat{E}_2}{8} \phi, \\
\bar{K} &= \frac{(\hat{E}_1 + \hat{E}_2)}{8} \phi.
\end{aligned} \qquad (30)$$

By setting $\hat{E}_1 = \hat{E}_2 = \hat{E}_0$ into Eqns. (29), it is immediate to recognize that the lattice response is isotropic with (stress-free) elastic moduli: $\bar{E} = \hat{E}_0 \ \phi/3$, $\bar{\nu} = 1/3$, $\bar{G} = \hat{E}_0 \ \phi/8$, and $\bar{K} = \hat{E}_0 \ \phi/4$ (cf. [9]).

It is useful to compare the above moduli with the Young modulus $\mathring{E}$, the Poisson's ratio $\mathring{\nu}$, thr shear modulus $\mathring{G}$, and the bulk modulus $\mathring{K}$, which are associated with the second-order constants given in Sect. 5.1.1. Upon replacing $\mathring{c}_{ij}$ and $\mathring{a}_{ij}$ respectively with $\mathring{C}_{ij}$ and the components $\mathring{A}_{ij}$ of the matrix $\mathring{A} = \mathring{C}^{-1}$, it is easy to verify that it results

$$\begin{aligned}
\mathring{E} &= \frac{(-E_1 \phi + 4\sigma_{01})(4\sigma_{01} + 4\sigma_{02} - (E_1 + E_2)\phi)}{2((2E_1 + E_2)\phi - 8\sigma_{01} - 4\sigma_{02})}, \\
\mathring{\nu} &= \frac{E_2 \phi - 4\sigma_{02}}{(2E_1 + E_2)\phi - 8\sigma_{01} - 4\sigma_{02}}, \\
\mathring{G} &= \frac{1}{8}(E_2 \phi - 4\sigma_{02}), \\
\mathring{K} &= K.
\end{aligned} \qquad (31)$$

Let us now numerically investigate on the values of the elastic stiffness moduli exhibited by a prestressed tetrakis lattice that employs either hard or soft FDM materials for the rods. Since for the lattice under consideration it results $V_S = 4A_1 h + 4A_2 h\sqrt{2}/2$, and we are assuming $A_2 = \sqrt{2}A_1$, from the definition of $\phi$ and Eqns. (24), we easily obtain

$$\begin{aligned}
A_1 &= \frac{V_0 \phi}{8h}, & \hat{\sigma}_1 &= \frac{\mathring{t}_1}{A_1} = \frac{4\sigma_{01}}{\phi}, & \hat{\varepsilon}_1 &= \frac{\hat{\sigma}_1}{\hat{E}_1} = \frac{4\sigma_{01}}{\hat{E}_1 \phi}, \\
A_2 &= \frac{V_0 \phi \sqrt{2}}{8h}, & \hat{\sigma}_2 &= \frac{\mathring{t}_2}{A_2} = \frac{4\sigma_{02}}{\phi}, & \hat{\varepsilon}_2 &= \frac{\hat{\sigma}_2}{\hat{E}_2} = \frac{4\sigma_{02}}{\hat{E}_2 \phi},
\end{aligned} \qquad (32)$$

where $\hat{\sigma}_1$ and $\hat{\varepsilon}_1$ denote the axial stress and axial strain carried by the rods of Type 1, while $\hat{\sigma}_2$ and $\hat{\varepsilon}_2$ denote the axial stress and axial strain carried by the Type 2 rods, respectively. Eqns. (24) highlight that the state of prestress $\sigma_{01}$ involves nonzero prestress forces only in the horizontal and vertical members (Type 1 rods), while $\sigma_{02}$ involves nonzero prestress forces only in the diagonal members (Type 2 rods). In order to prestress the lattice through an isotropic plane deformation from the natural state, we link $\sigma_{01}$ to $\sigma_{02}$ by imposing that it results $\hat{\varepsilon}_1 = \hat{\varepsilon}_2$ in Eqns. (32), which gives

$$\sigma_{02} = \frac{\hat{E}_2}{\hat{E}_1} \sigma_{01}. \tag{33}$$

Let $\hat{\sigma}_{y1}$ and $\hat{\sigma}_{y2}$ respectively denote the yield stresses of the materials that form the members of Type 1 and Type 2, and let us introduce the notations $\hat{\sigma} = \hat{\sigma}_1$, $\hat{\sigma}_y = min\ (\hat{\sigma}_{y1}, \hat{\sigma}_{y2}\hat{E}_1/\hat{E}_2)$. Figure 2 illustrates the distributions of the $E\ /\ \overline{E}$, $\nu\ /\ \overline{\nu}$, $G\ /\ \overline{G}$, and $K\ /\ \overline{G}$ ratios obtained via Eqns. (29)-(30), when the the dimensionless prestress variable $\hat{\sigma}/\hat{\sigma}_y$ ranges in the interval [0,1]. As we already noticed, we indeed focus our attention on lattices subject to equal biaxial tension, on considering that compressed lattices may suffer anticipated failure due to buckling [13, 14]. We use the acronyms 'Hard1&2', 'Soft1&2', 'Hard1/Soft2' and 'Soft1/Hard2' to respectively denote the lattices that employ ABS for the rods of Type 1 and 2, NJF for the rods of Type 1 and 2, ABS for the rods of Type 1 and NJF for the rods of Type 2, and NJF for the rods of Type 1 and ABS for the rods of Type 2. Table 1 provides the reference, stress-free moduli exhibited by the lattices under consideration. One observes from Tab. 1 that the Poisson's ratio of Hard1/Soft2 lattices is close to zero (as well as the corresponding shear modulus), while the value of such a property in Soft1/Hard2 lattices is very close to the extreme value (1) reachable by stable isotropic materials in 2d [50].

The plots in Fig. 2 highlight small variations of the elastic stiffness moduli with the prestress $\hat{\sigma}/\hat{\sigma}_y$ in homogeneous Hard lattices (few percents); large or moderately large variations in homogeneous Soft lattices; and markedly large variations in some composite Hard-Soft systems. Focusing our attention on systems that highlight more marked variations of the elastic stiffness moduli over the stress-free values, we note that $E\ /\ \overline{E}$ oscillates between 1.00 and 6.83 in Soft1/Hard2 composite lattices, while $\nu\ /\ \overline{\nu}$ oscillates between 1.00 and $-0.88$ in in homogeneous Soft lattices (being equal to zero for $\hat{\sigma}/\hat{\sigma}_y = 0.50$), and between 1.00 and $-5.04$ in Soft1/Hard2 composite lattices ($\nu = 0$ at $\hat{\sigma}/\hat{\sigma}_y \approx 0.17$). The $G\ /\ \overline{G}$ ratio ranges between 1.00 and 2.20 in homogeneous Soft lattices, and between 1.00 and 7.01 in Soft1/Hard2 composite lattices, always when the prestress ratio $\hat{\sigma}/\hat{\sigma}_y$ ranges between 0 and 1 (see Fig. 2). The variation of the shear modulus $G$ with the applied prestress is clearly understood by making use of (32)-(33) into Eqn. (29), obtaining

$$G = \frac{\hat{E}_2\ \phi}{8} + \left(2 + \frac{\hat{E}_2}{\hat{E}_1}\right)\frac{\hat{\sigma}\ \phi}{8}. \tag{34}$$

Eqn. (34) points out that $G$ might markedly change from the reference value $\overline{G} = \hat{E}_2\ \phi/8$, when $\hat{\sigma}$ can assume values close to $\hat{E}_2$, i.e. in composite systems that use a hard material for the rods of Type 1 and a soft material for the rods of Type 2 ($\hat{\sigma}_y = 2.5\hat{E}_2$). In particular, the difference $(G - \overline{G})$ assumes positive values under tensile isotropic prestress ($\hat{\sigma} > 0$), and negative values under isotropic pressure preload ($\hat{\sigma} < 0$). Large oscillations also exhibits the $\nu\ /\ \overline{\nu}$ ratio in Hard1/Soft2 composite systems, with the difference that such a quantity assume negative values (giving rise to *incremental auxetic response*) under tensile isotropic prestress, and positive values under isotropic pressure preload. Tab. 1 highlights that the reference values $\overline{G}$ and $\overline{\nu}$ of the above moduli assume very small values (close to zero) in correspondence to Hard1/Soft2 systems. By combining the results presented in Tab. 1 and Fig. 2, we conclude that in Soft1&2 lattices the elastic stiffness Poisson's ratio $\nu$ is equal to 0.33 for $\hat{\sigma} = 0$, while it results $\nu = -0,29$ for $\hat{\sigma} = \hat{\sigma}_y$ ($\nu = 0$ for $\hat{\sigma} = 0.50\ \hat{\sigma}_y$; $\nu = -0,15$ for $\hat{\sigma} = 0.75\ \hat{\sigma}_y$). In Hard1/Soft2 lattices such a quantity is instead equal to 0.002 for $\hat{\sigma} = 0$, and it results $\nu = -0,011$ for $\hat{\sigma} = \hat{\sigma}_y$ ($\nu = -0,004$ for $\hat{\sigma} = 0.5\ \hat{\sigma}_y$). A qualitatively different response is exhibited by the $E\ /\ \overline{E}$ ratio, which features large oscillations in composite systems that use a soft material for the rods of Type 1 and a hard material for the rods of Type 2, as we already observed (cf. Fig. 2). For what concerns the $K\ /\ \overline{K}$ ratio, we observe from Fig. 2 that such a quantity exhibits very small ($\sim 1\%$) deviations from one in full Hard systems, and oscillates between 1.00 and 0.60 in full Soft systems. The $K\ /\ \overline{K}$ ratio in composite

Hard1/Soft2 and Soft1/Hard2 systems (not shown in Fig. 2) remains very close to one, as in full Hard systems. Finally, it is easy to verify that the $\circ E / \overline{E}$, $\circ G / \overline{G}$, and $\circ K / \overline{K}$ ratios vary with $\hat{\sigma}/\hat{\sigma}_y$ exactly like the $K / \overline{K}$ ratio, while $\circ \nu / \overline{\nu}$ remains equal to one, independently of the applied prestress (cf. Eqns. (31)).

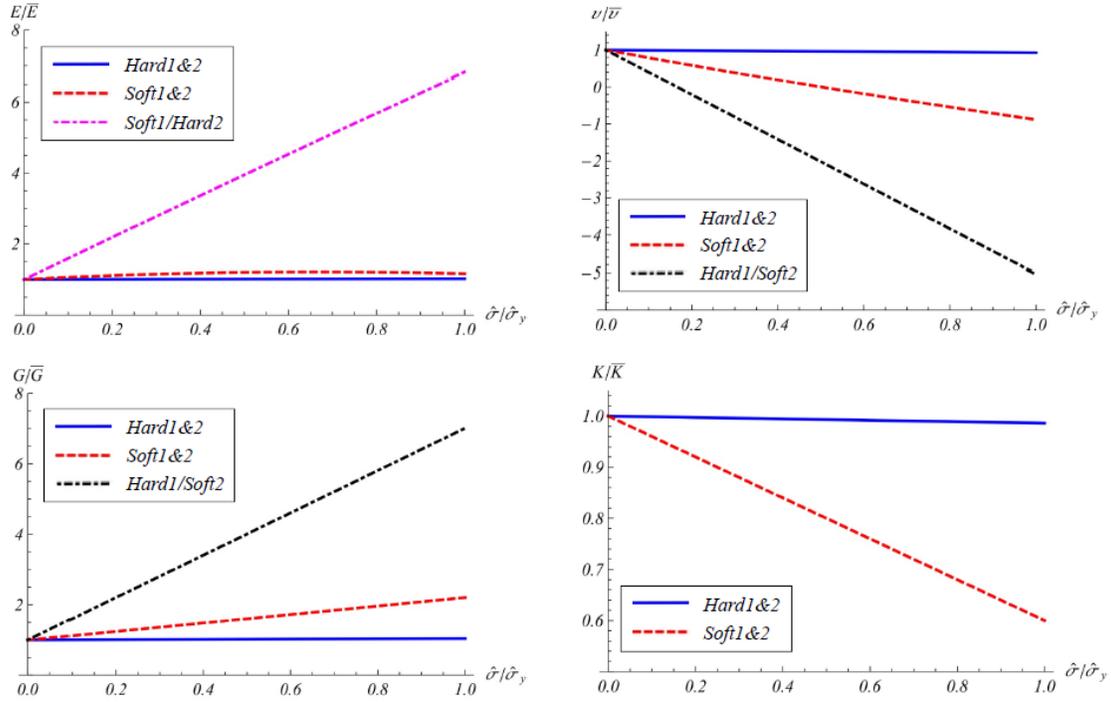

Figure 2: Elastic stiffness moduli of a tetrakis lattice material vs. the applied prestress for different arrangements of soft and hard rods.

Table 1: Elastic stiffness moduli in the stress-free state of tetrakis lattice materials

|  | $\overline{E}$ (MPa) | $\overline{\nu}$ | $\overline{G}$ (MPa) | $\overline{K}$ (MPa) |
|---|---|---|---|---|
| Hard1&2 | 0.7333 $\phi$ | 0.3333 | 0.2750 $\phi$ | 0.5500 $\phi$ |
| Soft1&2 | 0.0033 $\phi$ | 0.3333 | 0.00125 $\phi$ | 0.0025 $\phi$ |
| Hard1/Soft2 | 0.5513 $\phi$ | 0.0023 | 0.0013 $\phi$ | 0.2763 $\phi$ |
| Soft1/Hard2 | 0.0050 $\phi$ | 0.9910 | 0.2750 $\phi$ | 0.2763 $\phi$ |

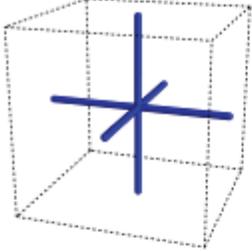

| | Unit vectors and initial forces |
|---|---|
| (cube diagram) | $\mathring{\mathbf{e}}_1, \ldots, \mathring{\mathbf{e}}_6 = \begin{bmatrix}\pm 1\\0\\0\end{bmatrix}, \begin{bmatrix}0\\\pm 1\\0\end{bmatrix}, \begin{bmatrix}0\\0\\\pm 1\end{bmatrix}$ |
| | $\mathring{t} = \dfrac{\sigma_0 V_0}{2\mathring{r}}$ |

Initial stress

$$\mathring{\sigma}_{\alpha\beta} = \delta_{\alpha\beta}\,\sigma_0 \qquad (\alpha,\beta = 1,2,3)$$

Second order elastic constants

$$\mathring{C}_{11} = \frac{\hat{E}_0\phi}{3} - \sigma_0$$
$$\mathring{C}_{12} = \mathring{C}_{44} = 0$$

Elastic stiffness constants

$$\mathring{c}_{11} = \mathring{C}_{11} + \sigma_0, \quad \mathring{c}_{12} = \mathring{C}_{12} - \sigma_0, \quad \mathring{c}_{44} = \mathring{C}_{44} + \sigma_0$$

Elastic stiffness moduli

$$E = \frac{\hat{E}_0^2\phi^2 - 3\hat{E}_0\phi\sigma_0 - 18\sigma_0^2}{3\hat{E}_0\phi - 9\sigma_0}$$

$$\nu = \frac{-27\sigma_0}{\hat{E}_0^3\phi^3 - 6\hat{E}_0^2\phi^2\sigma_0 - 9\hat{E}_0\phi\sigma_0^2 + 54\sigma_0^3}$$

$$G = \frac{\hat{E}_0\phi}{3}$$

$$K = \frac{1}{9}(E_0\,\phi - 6\,\sigma_0)$$

Table 2: Elastic constants of a simple cubic lattice material ($Z = 6$) under isotropic prestress.

### 5.2 3D lattices

The present section examines a variety of 3D lattice materials ($d = 3$) that match the equilibrium conditions (9) under arbitrary incremental deformations from the isotropically prestressed state. The examined lattices correspond to the simple cubic ($Z = 6$), body-centered cubic (BCC, $Z = 8$), face-centered cubic octet (FCC octet, $Z = 12$) and tetrakaidecahedral ($Z = 14$) truss structures analyzed in [10] under zero initial stress.

Tables 3-6 present the unit cell geometry and the 'prestressed' elastic constants obtained through the theory presented in Sect. 4. In the lattices with coordination numbers $Z = 6, 8, 12$ we observe that all the rods have the same length $\circ r$, cross section area A, and Young modulus $\hat{E}_0$, and carry equal initial (axial) forces $\circ t$. In the tetrakaidecahedral case ($Z = 14$), the lattice is instead made of six rods with length $\sqrt{2}h$, cross-section area $A_1$, Young modulus $\hat{E}_1$, and initial axial force $\circ t_1 = 4h\sigma_{01}$ (rods of 'Type 1' parallel to the cubic symmetry axes, represented in blue/dark color in Tab. 6); and eight rods with length $\sqrt{3/2}h$, cross-section area $A_2 = \eta A_1$, Young modulus $\hat{E}_2$, and initial force $\circ t_2 = 2\sqrt{3}h\sigma_{02}$ ('diagonal' rods of 'Type 2', represented in red/light color in Tab. 6). Here, $h$ denotes the lattice constant (length of the edges of the Kelvin-type unit cell of the lattice, cf. Tab. 6),

while $\sigma_{01}$ and $\sigma_{02}$ denote two different states of isotropic prestress, with the first one involving prestress forces only in the rods of Type 1, and the second one involving prestress forces only in the rods of Type 2.

It is easily shown that all the lattices in Tables 3-6 exibit cubic elastic symmetry, both in terms of second-order elastic constants $\circ C_{\alpha\beta\lambda\mu}$, and in terms of the elastic stiffness constants $\circ c_{\alpha\beta\lambda\mu}$, with the following independent constants different from zero (Voigt notation)

$$\mathring{C}_{11} = \mathring{C}_{1111} = \mathring{C}_{2222} = \mathring{C}_{3333}, \qquad \mathring{c}_{11} = \mathring{c}_{1111} = \mathring{c}_{2222} = \mathring{c}_{3333},$$
$$\mathring{C}_{12} = \mathring{C}_{1122} = \mathring{C}_{1133} = \mathring{C}_{2233}, \qquad \mathring{c}_{12} = \mathring{c}_{1122} = \mathring{c}_{1133} = \mathring{c}_{2233}, \tag{35}$$
$$\mathring{C}_{44} = \mathring{C}_{1212} = \mathring{C}_{1313} = \mathring{C}_{2323}, \qquad \mathring{c}_{44} = \mathring{c}_{1212} = \mathring{c}_{1313} = \mathring{c}_{2323}.$$

The results given in Tabs. 3-6 include the analytic expressions of the elastic stiffness moduli defined as follows

$$\begin{aligned}
E &= \frac{1}{\mathring{a}_{11}} = \frac{1}{\mathring{a}_{22}} = \frac{1}{\mathring{a}_{33}}, \\
\nu &= -\mathring{a}_{\alpha\beta} E \quad (\alpha,\beta = 1,2,3,\ \alpha \neq \beta), \\
G &= \frac{1}{\mathring{a}_{44}} = \frac{1}{\mathring{a}_{55}} = \frac{1}{\mathring{a}_{66}}, \\
K &= \frac{1}{3(\mathring{a}_{11} + 2\mathring{a}_{12})}.
\end{aligned} \tag{36}$$

where $\circ a_{ij}$ are the components of the compliance matrix $\circ a = \circ c^{-1}$. It is easy to verify that the results provided in Tabs. 3-6 exactly reduce to to those presented in [10, 9] when the applied prestress vanishes ($\sigma_0 = 0$).

In the case of the tetrakaidecahedral lattice, the elastic stiffness moduli (36) are provided for $\eta = A_2/A_1 = 3\sqrt{3}/4$ [9], when it results

$$\begin{aligned}
A_1 &= \frac{8h^2}{15}\phi, & \hat{\sigma}_1 &= \frac{15}{2}\frac{\sigma_{01}}{\phi}, & \hat{\varepsilon}_1 &= \frac{15}{2}\frac{\sigma_{01}}{\hat{E}_1\ \phi}, \\
A_2 &= \frac{2\sqrt{3}}{5}h^2\phi, & \hat{\sigma}_2 &= 5\frac{\sigma_{02}}{\phi}, & \hat{\varepsilon}_2 &= \frac{5}{\hat{E}_2}\frac{\sigma_{02}}{\phi}.
\end{aligned} \tag{37}$$

For such a lattice, the condition of volumetric deformation from the natural state leads to the following relationship between $\sigma_{01}$ and $\sigma_{02}$ (cf. Eqns. (37))

$$\sigma_{02} = \frac{3}{2}\frac{\hat{E}_2}{\hat{E}_1}\sigma_{01}. \tag{38}$$

By making use of the same notation introduced for tetrakis lattices in Sect. 5.1.3, we show in Fig. 8 the variation of the ratios between the elastic stiffness moduli $E$, $\nu$ and $G$ of a prestressed tetrakaidecahedral lattice and the stress-free moduli $\bar{E}$, $\bar{\nu}$ and $\bar{G}$, for varying values of the prestress variable $\hat{\sigma}/\hat{\sigma}_y$ in the interval [0,1] (as in Sect. 5.1.3, hereafter we use the subscripts 1 and 2 to mark properties relative to members of Type 1 and Type 2, respectively, and we set: $\hat{\sigma} = \hat{\sigma}_1$, $\hat{\sigma}_y = \min(\hat{\sigma}_{y1}, \hat{\sigma}_{y2}\hat{E}_1/\hat{E}_2)$). Table 7 provides the reference moduli exhibited by the tetrakaidecahedral lattices under consideration.

The plots illustrated in Fig. 8 provide results that are similar to those presented in Fig. 2 for a 2D tetrakis lattice, highligthing marked variations of the $E/\bar{E}$ ratio in composite lattices with soft rods of Type 1 and hard rods of Type 2 (Soft1/Hard2 systems), and marked variations of the $\nu/\bar{\nu}$ and $G/\bar{G}$ ratios in lattices composed of all soft rods (Soft systems), or lattices formed by hard rods of Type 1 and soft rods of Type 2 (Hard1/Soft2 systems). Also in tetrakaidecahedral lattices, as well as in tetrakis lattices, the Poisson ratio can assume negative values in Hard1/Soft2 systems under (isotropic) tensile preload, giving raise to incremental auxetic response from the prestressed state. A combined exam of the results presented in Tab. 7 and Fig. 8 reveals that in Soft1&2 tetrakaidecahedral lattices the elastic stiffness Poisson's ratio $v$ is equal to 0.25 for $\hat{\sigma} = 0$, while it results $v = -0,58$ for $\hat{\sigma} = \hat{\sigma}_y$ ($v = -0.06$ for $\hat{\sigma} = 0.50 \ \hat{\sigma}_y$; $v = -0,29$ for $\hat{\sigma} = 0.75 \ \hat{\sigma}_y$). In Hard1/Soft2 lattices such a quantity is instead equal to 0.002 for $\hat{\sigma} = 0$, and it results $v - 0,012$ for $\hat{\sigma} = \hat{\sigma}_y$ ($v = -0,005$ for $\hat{\sigma} = 0.5 \ \hat{\sigma}_y$). The plots in Fig. 8 also show that the $\overset{\circ}{K}/\overline{K}$ ratio exhibits very small ($\sim 1\%$) deviations from one in full Hard systems, and that the same ratio oscillates between 1.00 and 0.20 in full Soft systems, when the prestress ratio $\hat{\sigma}/\hat{\sigma}_y$ ranges between 0 and 1 (cf. Fig. 8).

We now pass to compute the Young modulus $\circ E$, Poisson's ratio $\circ v$, shear modulus $\circ G$, and bulk modulus $\circ K$ associated with the second-order constants $\circ C_{ij}$. It is an easy task to verify that such quantities assume the following expressions in the case of the tetrakaidecahedral lattices under consideration

$$\overset{\circ}{E} = \frac{(2 E_1 \phi - 15 \sigma_{01})(2 E_1 \phi + 3 E_2 \phi - 15 \sigma_{01} - 15 \sigma_{02})}{15 \left\{ 2 \left[ (E_1 + E_2) \phi + 5 \sigma_{02} \right] - 15 \sigma_{01} \right\}},$$

$$\overset{\circ}{v} = \frac{E_2 \phi - 5 \sigma_{02}}{2 \left[ (E_1 + E_2) - 5 \sigma_{02} \right]},$$

$$\overset{\circ}{G} = \frac{1}{15} (E_2 \phi - 5 \sigma_{02}),$$

$$\overset{\circ}{K} = \frac{1}{45} (2 E_1 \phi + 3 E_2 \phi - 15 \sigma_{01} - 15 \sigma_{02}) \tag{39}$$

Eqns. (39) highlight that it results $\circ K = K + (\sigma_{01} + \sigma_{02})/3$ in the present 3D case, wile in 2D one gets $\circ K = K$ (cf. Sect. 5.1). It is easily shown that all the $\circ E / \overline{E}$, $\circ G / \overline{G}$, and $\circ K / \overline{K}$ ratios exhibit the same variation law with the prestress $\hat{\sigma}/\hat{\sigma}_y$, featuring $\sim 1\%$ deviations from one in full Hard systems, and oscillations between 1.00 and 0.60 in full Soft systems (Fig. 9). The $\circ v / \overline{v}$ ratio is instead constantly equal to one, for any $\hat{\sigma}/\hat{\sigma}_y \in [0,1]$, as in the case of the tetrakis lattices analyzed in Sect. 5.1.

| | |
|---|---|
| 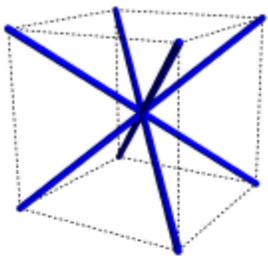 | Unit vectors and initial forces $$\mathring{e}_1, \ldots, \mathring{e}_8 = \begin{bmatrix} \pm 1/\sqrt{3} \\ \pm 1/\sqrt{3} \\ \pm 1/\sqrt{3} \end{bmatrix}$$ $$\mathring{t} = \frac{3\sigma_0 V_0}{8\mathring{r}}$$ |

**Initial stress**

$$\mathring{\sigma}_{\alpha\beta} = \delta_{\alpha\beta}\sigma_0 \qquad (\alpha,\beta = 1,2,3)$$

**Second order elastic constants**

$$\mathring{C}_{11} = \mathring{C}_{12} = \mathring{C}_{44} = \frac{\hat{E}_0\phi - 3\sigma_0}{9}$$

**Elastic stiffness constants**

$$\mathring{c}_{11} = \mathring{C}_{11} + \sigma_0, \quad \mathring{c}_{12} = \mathring{C}_{12} - \sigma_0, \quad \mathring{c}_{44} = \mathring{C}_{44} + \sigma_0$$

**Elastic stiffness moduli**

$$E = \frac{3\left(\hat{E}_0\phi - 6\sigma_0\right)\sigma_0}{\hat{E}_0\phi - 3\sigma_0}$$

$$\nu = \frac{\hat{E}_0\phi - 12\sigma_0}{2\hat{E}_0\phi - 6\sigma_0}$$

$$G = \frac{\hat{E}_0\phi + 6\sigma_0}{9}$$

$$K = \frac{1}{9}(E_0\phi - 6\sigma_0)$$

Table 3: Elastic constants of a BCC lattice material ($Z = 8$) under isotropic prestress.

| | Unit vectors and initial forces |
|---|---|
| 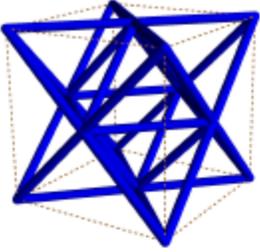 | $\mathring{\mathbf{e}}_1, ..., \mathring{\mathbf{e}}_{12} = \begin{bmatrix} 0 \\ \pm 1/\sqrt{2} \\ \pm 1/\sqrt{2} \end{bmatrix}, \begin{bmatrix} \pm 1/\sqrt{2} \\ 0 \\ \pm 1/\sqrt{2} \end{bmatrix}, \begin{bmatrix} \pm 1/\sqrt{2} \\ \pm 1/\sqrt{2} \\ 0 \end{bmatrix}$ $\mathring{t} = \dfrac{\sigma_0 V_0}{4\mathring{r}}$ |

**Initial stress**

$\mathring{\sigma}_{\alpha\beta} = \delta_{\alpha\beta}\sigma_0 \qquad (\alpha,\beta = 1,2,3)$

**Second order elastic constants**

$\mathring{C}_{11} = \dfrac{\hat{E}_0\phi - 3\sigma_0}{6}$

$\mathring{C}_{12} = \mathring{C}_{44} = \dfrac{\hat{E}_0\phi - 3\sigma_0}{12}$

**Elastic stiffness constants**

$\mathring{c}_{11} = \mathring{C}_{11} + \sigma_0, \; \mathring{c}_{12} = \mathring{C}_{12} - \sigma_0, \; \mathring{c}_{44} = \mathring{C}_{44} + \sigma_0$

**Elastic stiffness moduli**

$E = \dfrac{\left(\hat{E}_0^2\phi^2 + 15\hat{E}_0\phi\sigma_0 - 126\sigma_0^2\right)}{9\hat{E}_0\phi - 27\sigma_0}$

$\nu = \dfrac{\hat{E}_0\phi - 15\sigma_0}{3\hat{E}_0\phi - 9\sigma_0}$

$G_a = \dfrac{\hat{E}_0\phi + 9\sigma_0}{12}$

$K = \dfrac{1}{9}(\hat{E}_0\phi - 6\sigma_0)$

Table 4: Elastic constants of a FCC lattice material ($Z = 12$) under isotropic prestress.

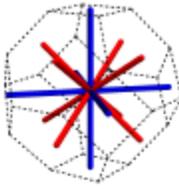

Table 5: Elastic constants of a tetrakaidecahedral lattice material ($Z = 14$) under isotropic prestress.

Table 6: Elastic stiffness moduli in the stress-free state of tetrakaidecahedral lattice materials

|  | $\bar{E}$ (MPa) | $\bar{v}$ | $\bar{G}$ (MPa) | $\bar{K}$ (MPa) |
| --- | --- | --- | --- | --- |
| Hard1&2 | 0.3667 $\phi$ | 0.2500 | 0.2750 $\phi$ | 0.2444 $\phi$ |
| Soft1&2 | 0.0017 $\phi$ | 0.2500 | 0.1467 $\phi$ | 0.0011 $\phi$ |
| Hard1/Soft2 | 0.2940 $\phi$ | 0.0023 | 0.0007 $\phi$ | 0.0984 $\phi$ |
| Soft1/Hard2 | 0.0020 $\phi$ | 0.4978 | 0.1467 $\phi$ | 0.1471 $\phi$ |

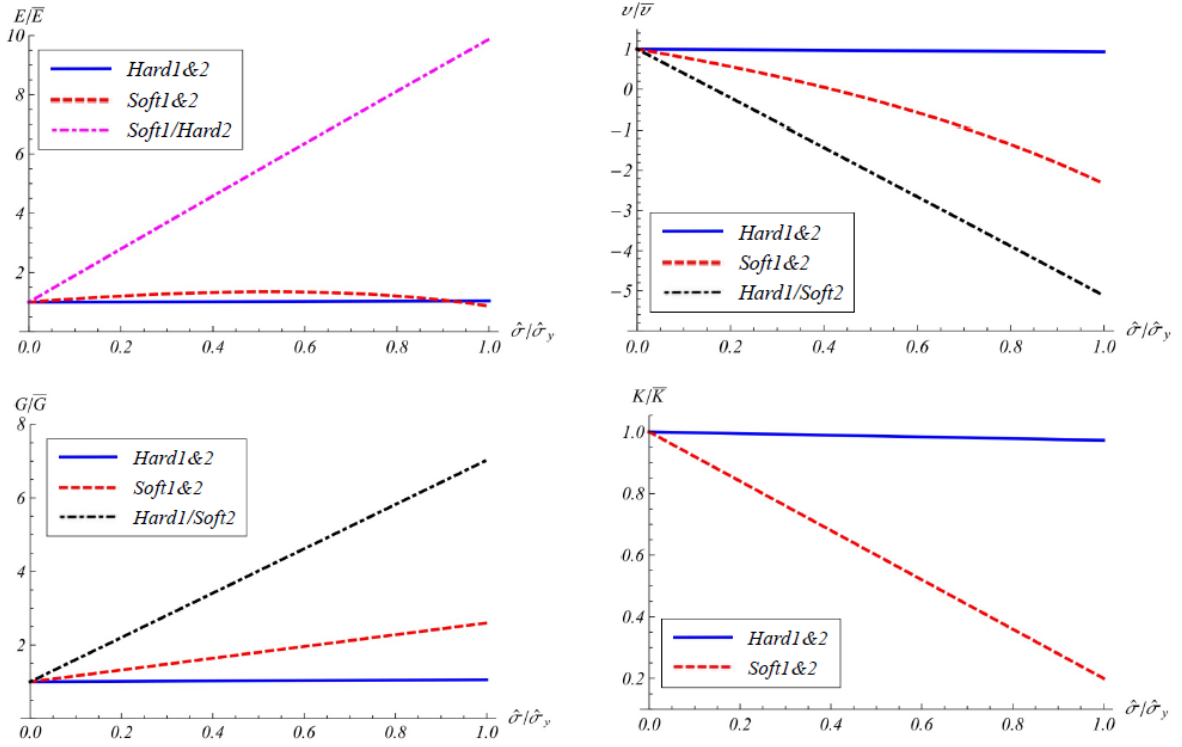

Figure 3: Elastic stiffness moduli of a tetrakaidecahedral lattice material vs. the applied prestress for different arrangements of soft and hard rods.

## 6  Concluding remarks

We have investigated the incremental elastic moduli of 2d and 3d stretching-dominated lattice materials from an isotropically prestressed state. The results presented in Sect. 5 have examined the variation with the applied prestress of the elastic stiffness moduli of a variety of lattices, with special focus on composite lattices composed of hard and/or soft FDM materials in different rods, and tensile preload. We have observed that the ratios between the incremental Young and shear moduli in the prestressed state and the corresponding stress-free moduli of composite lattices may reach extreme values equal to $7 \div 8$, giving rise to significant increases of the tangent stiffness. On the other hand, we have noted that the    elastic stiffness Poisson's ratio may show markedly negative values in the soft and hard/soft lattices analyzed in in Sect. 5, by reaching extreme values    that can be equal to $\approx -2 \div -5$ times the value in the stress-free state, under biaxial/triaxial tension preloads. Such a result reveals a novel feature of isotropically prestressed lattices, which consists of the possibility to enforce and modulate an incremental auxetic response from the prestressed state, through a suitable control of the tensile preload.

The state of prestress examined in the present work can be, e.g., applied by constraining a lattice structure to reaction walls, like, e.g., in the case of sandwich structures composed of stiff facesheets and truss cores [44]. Actuated connections and/or truss members can be employed to enforce the desired prestress [37, 45]. We address the fabrication and testing of the physical models of the structures anaylzed in this study to future work, through recourse to traditional fabrication methods or novel additive manufacturing techniques. The latter may, e.g., employ projection micro-stereolitography and swelling materials [42, 43], and/or multi-jet technologies that are able to simultaneously 3d print materials with different coefficients of thermal expansion. Additional future research lines will address the generalization of the approach presented in this study to different states of initial stress and lattice geometries.

**Acknowledgements**
A.A. gratefully acknowledges financial support from the Department of Civil Engineering at the University of Salerno. The authors wish to thank Enrico Babilio (University of Naples "Federico II") and Marc Durand (University Paris Diderot) for many useful discussions.

**Compliance with Ethical Standards**
The authors declare that they have no conflict of interest.